# Extending the scanning angle of a phased array antenna by using a null-space medium

F. Sun [1, 2] and S. He [1, 2]

1 Centre for Optical and Electromagnetic Research, Zhejiang Provincial Key Laboratory for Sensing Technologies, China
2 Department of Electromagnetic Engineering, School of Electrical Engineering, Royal Institute of Technology (KTH), Sweden

**Abstract**
By introducing a columnar null-space region as the reference space, we design a radome that can extend the scanning angle of a phased array antenna (PAA) by a predetermined relationship (e.g. a linear relationship between the incident angle and steered output angle can be achieved). After some approximation, we only need two homogeneous materials to construct the proposed radome layer by layer. This kind of medium is called a null-space medium, which has been studied and fabricated for realizing hyper-lenses and some other devices. Numerical simulations verify the performance of our radome.

**Introduction**
The phased array antenna (PAA) plays an important role in radar, broadcast, communication, etc. [1, 2]. In practice, due to various factors, the scanning range of a PAA is often within $-60^o$ to $+60^o$ [1, 2]. It is still extremely challenging to achieve a scanning angle larger than $60^o$ with an acceptable beam quality. One applicable method is to set a radome behind the PAA to extend the angle of the output beam produced by PAA. *Lam et al.* has designed and fabricated a radome that can extend the scanning angle of a PAA to almost $-90^o$ to $+90^o$ [3, 4]. However, the proposed radome, according to heavy use of simulation optimization, cannot achieve a linear relationship between the input beam and the output beam, which is often desired in practice.

  Recently, we proposed a general method to design the radome that can extend the scanning angel of a PAA by using transformation optics (TO) [5]. The relationship between the input angle and the output angle of our radome can be manipulated as required (e.g. a linear relationship can be achieved). However, the material of the radome proposed in reference [5] is very hard to be acquired experimentally: it requires an inhomogeneous anisotropic medium (see Eq. (8) in [5]). In this paper, we will introduce a columnar null-space region as the reference space (but not free space as in [5]) to design a novel radome that keeps the advantage of our previous design (a linear relationship between input and output beams can be achieved) and only requires two homogeneous media to realize it layer by layer. The main features of the proposed radome in this paper include: 1) it can be used to extend the scanning range of a PAA (e.g. achieve a scanning angle larger than $60^o$); 2) a predetermined linear relationship between the input and output beams can be achieved; and 3) the material requirement of the proposed radome is much simper than our previous design [5]. Only two layered homogeneous anisotropic media can be used to construct the proposed radome. Our methods in this paper will further advance the practical realization of this high quality radome with predetermined steering functionality.

**Method and Simulation**

Our design is based on TO [6, 7] and a null-space medium [8, 9]. TO is based on the form-invariant of Maxwell's equations under coordinate transformation. A special medium (the so-called transformed medium), which can be used to control the tracing of electromagnetic waves [6, 7] or the distribution of DC electric or magnetic fields [10-15] in an unprecedented way, can be designed by TO. This idea has been extended to control the tracing of acoustic waves [16] and thermal fields [17, 18]. The null-space medium [8, 9] is also designed with transformation optics. We have previously used the idea of further transforming inside a null-space medium to reduce the material requirement of a DC magnetic lens [13]. TO can help to establish a relationship between two spaces: one is the reference space and the other is the real space. By designing the coordinate transformation between these two spaces, we can obtain the transformed medium in the real space with a pre-designed function.

We have previously provided a general method to design a radome of arbitrary shape that can extend the scanning angle of a PAA [5]. However, as the reference space in that paper is free space, the material requirement of the designed radome in [5] is very complex. In this paper, we will introduce the null-space medium as the reference space. In order to describe our idea analytically, we choose a columnar radome in our design. First we will introduce the idea of a columnar null-space region. Assuming the reference space is free space, we perform the following transformation:

$$r' = \begin{cases} \dfrac{R_1}{R_2 - \Delta} r, & r \in [0, R_2 - \Delta] \\ \dfrac{d}{\Delta} r - \dfrac{d - \Delta}{\Delta} R_2, & r \in [R_2 - \Delta, R_2], \quad \theta' = \theta, \ z' = z, \\ r, & r \in [R_2, \infty) \end{cases} \quad (1)$$

where $d = R_2 - R_1$, $0 < \Delta < R_1 < R_2$. The quantities in the reference or real space (before or after the transformation) is expressed without or with primes, respectively. If the reference space is filled with the free space, we can obtain the transformed medium in the real space after the coordinate transformation (1) [6, 7]:

$$\varepsilon'_{cy} = \mu'_{cy} = \begin{cases} diag\left(1,1,\left(\dfrac{R_2 - \Delta}{R_1}\right)^2\right) \xrightarrow{\Delta \to 0} diag\left(1,1,\left(\dfrac{R_2}{R_1}\right)^2\right), & r' \in [0, R_1] \\ diag\left(P, \dfrac{1}{P}, \left(\dfrac{\Delta}{d}\right)^2 P\right) \xrightarrow{\Delta \to 0} diag(\infty, 0, 0), & r' \in [R_1, R_2] \\ 1, & r' \in [R_2, \infty) \end{cases} \quad (2)$$

where $P = 1 + (d / \Delta - 1) R_2 / r'$. The subscript 'cy' means the quantity is expressed in the cylindrical coordinate system. If $\Delta$ approaches zero, the region $R_2 - \Delta < r < R_2$ in the reference space will approach a cylindrical surface and not a space. In this case, the region of space $R_1 < r' < R_2$ in the real space also corresponds to a surface and not a space. Therefore we refer to the medium in the region $R_1 < r' < R_2$ in the real space as the null-space medium. It should be noted that we have to introduce an anisotropic medium into the region $r' < R_1$ in order to maintain a continuous transformation. The columnar null-space medium in Eq. (2) has been studied in other literatures [8, 9], e.g., a hyper-lens that can provide high-resolution imaging is a columnar null-space medium.

In this paper, we will use the columnar null-space medium in Eq. (2) as the reference space, and then make a further transformation inside this null-space medium to achieve a radome that can extend the scanning angle of a PAA. We express the coordinates in the reference space and the real space as $(x', y', z')$ and $(x'', y'', z'')$, respectively. The medium in the reference space is the columnar null-space medium described by Eq. (2). We choose the following transformation:

$$r'' = r';$$
$$\theta'' = \begin{cases} \theta', & r' \in [0, R_1] \\ f(\theta', r'), & r' \in [R_1, R_2] \\ \theta', & r' \in [R_2, \infty) \end{cases} \quad (3)$$
$$z'' = z'$$

where $f$ is a continuous function, which determines the relationship between the input angle and output angle of the radome [5]. In this paper, we only consider the simplest relation between the input and output angles of the radome (an angle shifter in [5]), which requires that $f$ satisfies the following boundary conditions:

$$\theta'' = f(\theta', r' = R_1) = \theta'$$
$$\theta'' = f(\theta', r' = R_2) = \theta' + \theta_0 \quad (4)$$

Eq. (4) indicates that we choose an identical transformation at the input surface of our radome $r'' = r' = R_1$ and a linear angle shift at the output surface of our radome $r'' = r' = R_2$. Therefore, the radome designed by Eq. (3) and (4) performs like a linear angle shifter: the angle of the output beam is the angle of the input beam shifted by a fixed angle $\theta_0$. By using TO [5, 6], we can obtain the transformed medium of the transformation described in Eqs. (3) and (4) when the reference space is expressed by Eq. (2):

$$\varepsilon''_{cy} = \mu''_{cy} = \begin{cases} diag(1,1,\left(\dfrac{R_2-\Delta}{R_1}\right)^2), & r'' \in [0, R_1] \\ \dfrac{P}{f_{\theta'}} \begin{bmatrix} 1 & r''f_{r'} & 0 \\ r''f_{r'} & (r''f_{r'})^2 + \dfrac{1}{P^2}(f_{\theta'})^2 & 0 \\ 0 & 0 & \left(\dfrac{\Delta}{d}\right)^2 \end{bmatrix}, & r'' \in [R_1, R_2] \quad (5) \\ 1, & r'' \in [R_2, \infty) \end{cases}$$

where $f_{r'} = \partial f / \partial r'$ and $f_{\theta'} = \partial f / \partial \theta'$. Note that the region $R_1 < r'' < R_2$ in Eq. (5) is our radome, and the PAA is placed in the region $r'' < R_1$. The region outside the radome $r'' > R_2$ is free space, in accordance with the given situation. If we define the null-space as the reference space, which means $\Delta$ approaches zero ($P$ also approaches zero), the region $R_1 < r'' < R_2$ in Eq. (5) can be approximately expressed as (we drop the infinitesimal of higher orders):

$$\varepsilon''_{cy} = \mu''_{cy} \approx \dfrac{P}{f_{\theta'}} \begin{bmatrix} 1 & r''f_{r'} & 0 \\ r''f_{r'} & (r''f_{r'})^2 & 0 \\ 0 & 0 & \left(\dfrac{\Delta}{d}\right)^2 \end{bmatrix}, r'' \in [R_1, R_2] \quad (6)$$

The expression in Eq. (6) is still in the cylindrical coordinate system. We can make a diagonalization to express it under the principal axis system:

$$\varepsilon''_p = \mu''_p = diag(\dfrac{P}{f_{\theta'}}(1+(r''f_{r'})^2), 0, \dfrac{P}{f_{\theta'}}\left(\dfrac{\Delta}{d}\right)^2) \xrightarrow[P=1+\left(\frac{d}{\Delta}-1\right)\frac{R_2}{r'}]{\Delta \to 0} diag(\infty, 0, 0), r'' \in [R_1, R_2] \quad (7)$$

The subscript 'p' means that the quantity is expressed in its principal axis system. As we can see from Eq. (7), the radome can be approximately treated as a highly anisotropic homogeneous medium when $\Delta$ approaches zero. The direction of the local pricipal axis system can be obtained by rotating an angle $\phi$ from the cylindrical coordinate system [11]:

$$\tan 2\phi = \dfrac{2(r''f_{r'})}{1-(r''f_{r'})^2} \Rightarrow \tan \phi = r''f_{r'} \quad (8)$$

As we can see from Eq. (8), $\phi$ is related with function $f$. Considering that the function $f$ can be any arbitrary continuous function that satisfies the boundary condition (4), $\phi$ can thus also be arbitrary accordingly. To summarize the parameters of the whole structure: if $\Delta$ approaches zero (we use the null-space medium as the

reference space), the whole structure described by Eq. (5) can be reduced to:

$$\begin{aligned}
\varepsilon''_{cy} &= \mu''_{cy} = diag(1,1,\left(\frac{R_2}{R_1}\right)^2), & r'' &\in [0, R_1] \\
\varepsilon''_p &= \mu''_p = diag(\infty, 0, 0), & r'' &\in [R_1, R_2] \\
\varepsilon''_{cy} &= \mu''_{cy} = 1, & r'' &\in [R_2, \infty)
\end{aligned} \quad (9)$$

We should note that although we can simplify the materials in the radome region $R_1 < r'' < R_2$ by introducing the null-space medium into the reference space, it also causes the region $r'' < R_1$ to be some anisotropic homogeneous medium rather than air.

Since the material of our radome in its principle axis system is a highly anisotropic homogeneous one (see Eq. (7)), we can use two isotropic media with extreme parameters to realize it layer by layer. According to the effective medium theory, the effective permittivity (or permeability) of two isotropic homogeneous media with $\varepsilon_1$ and $\varepsilon_2$ (or $\mu_1$ and $\mu_2$) can be given as:

$$\begin{aligned}
\varepsilon_\|^{-1} &= \varepsilon_1^{-1} f_1 + \varepsilon_2^{-1} f_2 \\
\varepsilon_\perp &= \varepsilon_1 f_1 + \varepsilon_2 f_2 \\
\mu_\|^{-1} &= \mu_1^{-1} f_1 + \mu_2^{-1} f_2 \\
\mu_\perp &= \mu_1 f_1 + \mu_2 f_2
\end{aligned} \quad (10)$$

The subscript $\|$ indicates the direction along the normal vector of the interface between two media and $\perp$ indicates the direction perpendicular to the normal vector of the interface between two media. $f_1$ and $f_2$ correspond to the filling factors of two media ($f_1 + f_2 = 1$). We can choose $\varepsilon_1$ and $\mu_1$ to be very large; $\varepsilon_2$ and $\mu_2$ become nearly zero to mimic a highly anisotropic medium, according to Eq. (9). In this case, the direction (angle $\phi$) of the principal axis in Eq. (8) is along the direction of the interface between two media. As $f$ can be an arbitrary continuous function, $\phi$ can be arbitrarily and continuously changing inside the radome. The only thing we need to note is that it should satisfy boundary condition (4), which requires the angle shift of angle $\alpha_1$ (the rotation angle of intersection between the input surface of the radome and the interface of the two media with respect to the $x''$-axis) and $\alpha_2$ (the rotation angle of intersection between the output surface of the radome and the interface of the two media with respect to the $x''$-axis) to be exactly $\theta_0$.

The schematic diagram of our radome can be shown in Fig. 1(a): the whole radome is composed of two isotropic media layer by layer (the materials requirement is greatly simplified compared with the former material requirement in [5]). One medium requires high permittivity and permeability, and the other medium requires both the permittivity and permeability to be nearly zero. High permittivity can be simply achieved by using metal (metal performs as a perfect conductor in the microwave band). High permeability can be achieved through the use of artificial perfect magnetic conductors [19]. The medium with nearly zero permittivity and permeability can be achieved with impedance-matched zero-index metamaterials [20].

Some literatures have studied how to realize such a null-space medium [8, 21].

Next we will utilize finite element method (FEM) to verify the performance of our radome. The FEM we use is based on the commercial software COMSOL Multiphysics. We consider a 2D TM wave case, and the two homogenous materials are chosen as:

$$\varepsilon_1 = 10^{-3}; \mu_1 = 10^{-3}; \varepsilon_2 = 10^{3}; \mu_2 = 10^{-3}. \qquad (11)$$

Fig. 2 shows the performance of the proposed radome. As we can see from Fig. 2(a) and (b), this null-space medium (composed of two homogeneous media in Eq. (11)) performs like a special waveguide: it confines the incident beam inside it, and bends the light beam as itself is bending. At the output surface of the radome, this bending waveguide rotates exactly an angle $\theta_0$, which leads to an angle shift $\theta_0$ between the input beam and the output beam. Strictly speaking, in order to make the layered composite material perform as an effective medium for electromagnetic waves, the period of these two materials should be much smaller than the wavelength of the electromagnetic wave (e.g. less than $\lambda / 10$). However, the computational domain in our simulation is very large (larger than $10\lambda$): if the period of these two materials is too small (e.g. less than $\lambda / 4$), the computation will exceed the memory limit of the computer. Therefore we have to set the period to be about $\lambda / 3$ in our simulation, which leads to some reflections at the input surface of our radome.

We also simulate the case where we replace the anisotropic medium in the region $r'' < R_1$ (see Eq. (9)) with air. As shown in Fig. 2(c) and (d), the radome still works in this case (note that the output angle is slightly affected: see Fig. 3).

For a 2D TE wave case, the two homogenous materials are chosen as:

$$\varepsilon_1 = 10^{-3}; \mu_1 = 10^{-3}; \varepsilon_2 = 10^{-3}; \mu_2 = 10^{3}. \qquad (12)$$

Simulation results verify the performance of radome composed by the two materials described by Eq. (12) (see Fig. 4).

**Discussion and Conclusion**

Based on transformation within a null-space medium, we propose a radome design that can extend the scanning angle of a PAA. It only requires two homogeneous materials (null-space media) to construct the proposed radome. Simulation results indicate good performance of our radome. Our work in this paper follows our previous work on how to design a high-performance radome that can extend the scanning angle of a PAA with a predetermined way, simplifies the material requirement of the proposed radome, and advances closer to practical realization.

**Acknowledgement**
This work is partially supported by the National High Technology Research and Development Program (863 Program) of China (No. 2012AA030402), the National Natural Science Foundation of China (Nos. 61178062 and 60990322), the Program of Zhejiang Leading Team of Science and Technology Innovation, Swedish VR grant (#

621-2011-4620) and SOARD. Fei Sun thanks the China Scholarship Council (CSC) NO. 201206320083.

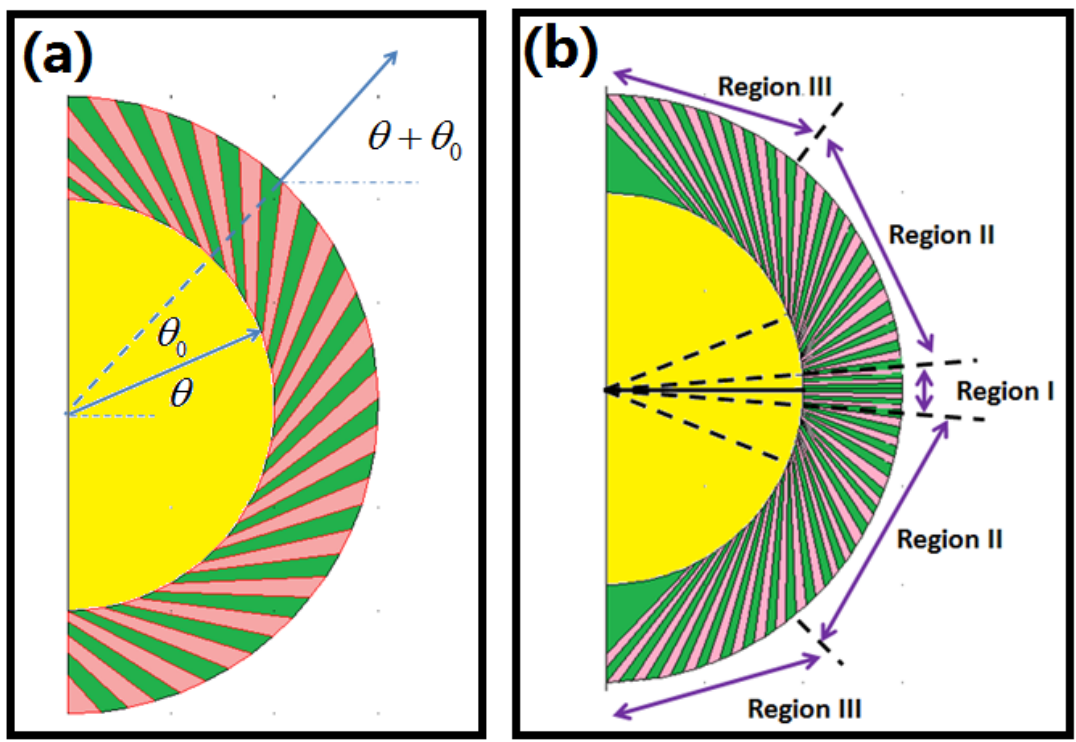

**Figure 1** (a) The basic structure of our radome, which can shift the incident beam by a fixed angle $\theta_0$. The radome is composed of two isotropic materials layer by layer: the permittivity and permeability of the green material is extremely high, and the permittivity and permeability of the red material is nearly zero. The yellow region ($r'' < R_1$) is the anisotropic medium. The white region is air. The materials distribution of the whole structure can be described by Eq. (9). (b) A multilevel radome that more closely resembles practical applications. This radome is divided into three different regions: region I with $\theta_0 = 0°$, which indicates that the incident beam will not be steered in this region; region III with a fixed shifting angle $\theta_0$, which is the same as (a) except that the lower part of region III is the mirror image of the upper part about the $x''$-axis; and region II, a transition region in which $\theta_0$ gradually changes from $0°$ to the fixed angle in region III.

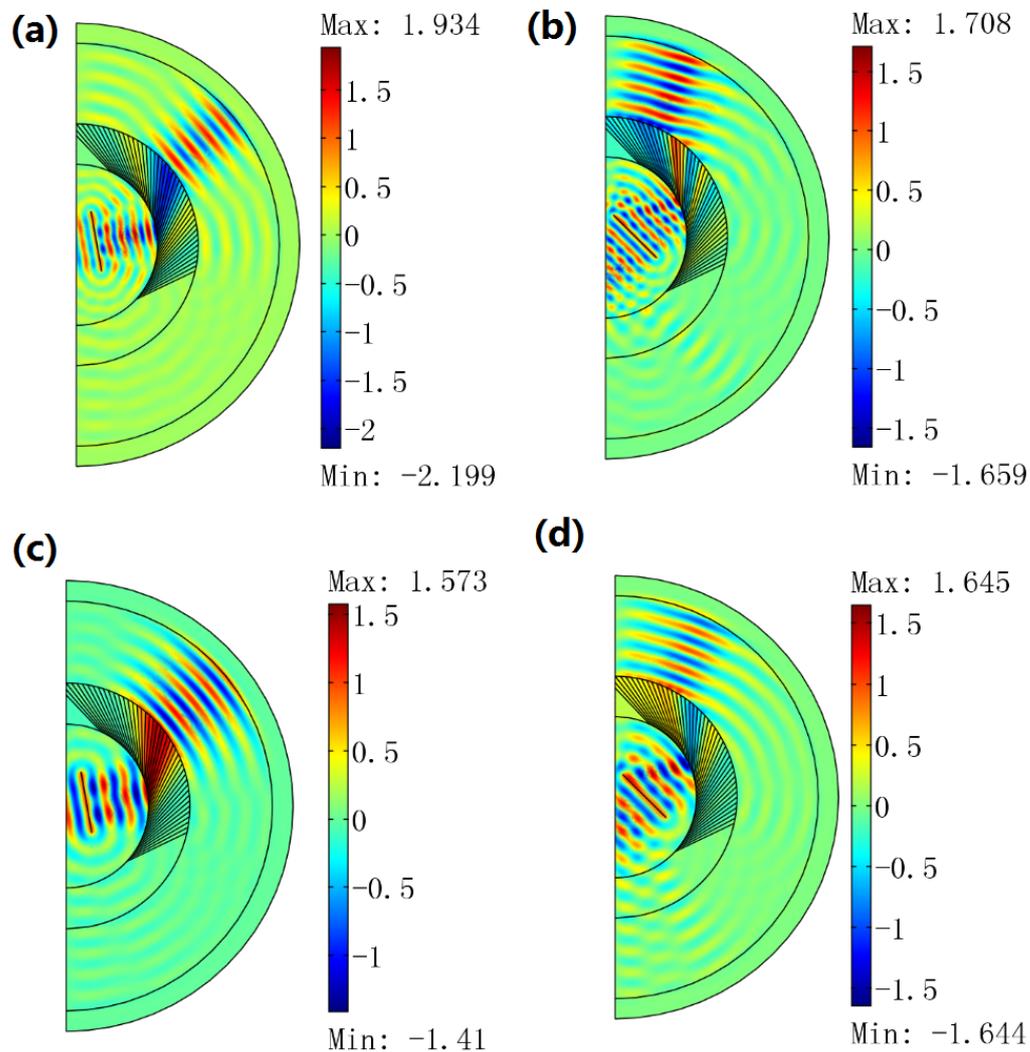

**Figure 2** 2D TM wave FEM simulation results: the $z$-component of the magnetic field. (a) and (c): a beam with angle $10°$ incident onto the radome, which is composed of the two materials described by Eq. (11). The shift angle of their interface relative to the $x''$-axis is $\alpha_2 - \alpha_1 = 30°$. The angles of the output beam are $39.9°$ and $39.5°$, respectively. (b) and (d): a beam with angle $45°$ incident onto the radome, which is composed of the two materials described by Eq. (11). The shift angle of their interface relative to the $x''$-axis is $\alpha_2 - \alpha_1 = 30°$. The angle of the output beam is $74°$ and $70°$, respectively. (a) and (b): the medium in the region $[0, R_1]$ is described by Eq. (9). (c) and (d): the medium in the region $[0, R_1]$ is air. To conserve memory during computation, we drop the lower part of the radome, which doesn't affect the results.

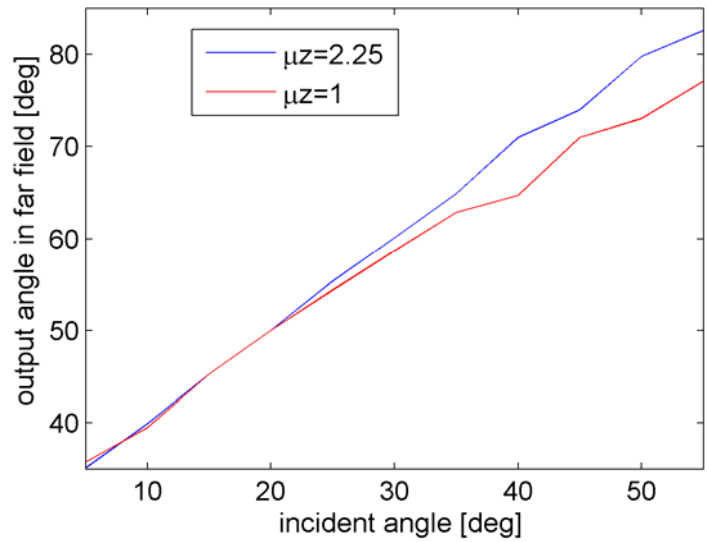

**Figure 3** 2D TM wave FEM simulation results: the relation between the angle of the input beam and the angle of the output beam in the far field for the radome composed by the two materials described in Eq. (11). The shift angle of their interface with respect to the *x''*-axis is $\alpha_2 - \alpha_1 = 30°$.

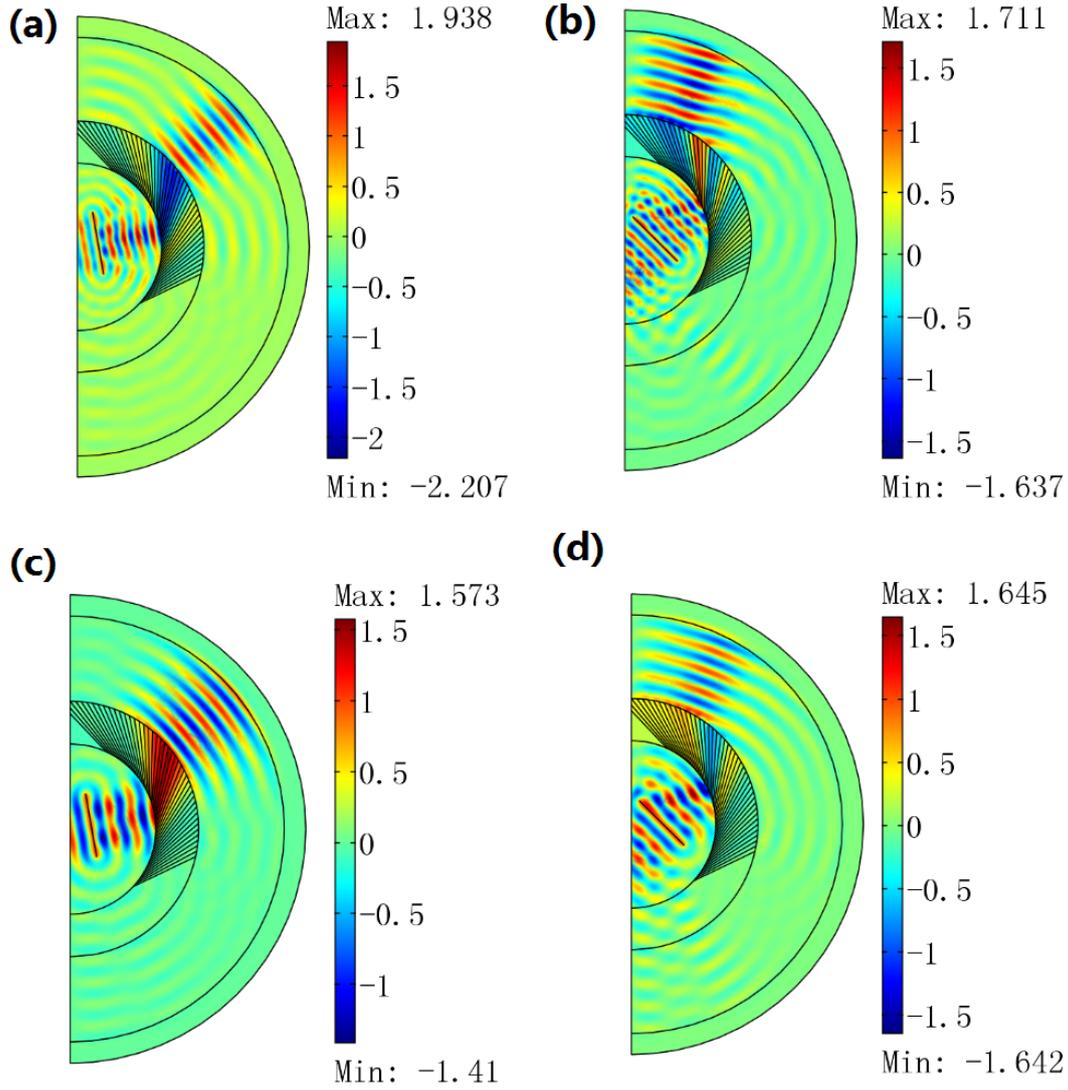

**Figure 4** 2D TE wave FEM simulation results: the *z*-component of the electric field. (a) and (c): a beam with angle 10° incident onto the radome (composed of the two materials described by Eq. (12)) with the shift angle of interface with respect to the *x"*-axis set to 30°. The angles of the output beam are 39.9° and 39.7°, respectively. (b) and (d): a beam with angle 45° incident onto the radome (composed of the two materials described by Eq. (12)) with the shift angle of interface with respect to the *x"*-axis set to 30°. The angles of the output beam are 74.2° and 70.2°, respectively. (a) and (b): the medium in the region [0, $R_1$] is described by Eq. (9). (c) and (d): the medium in the region [0, $R_1$] is air. To conserve memory during computation, we drop the lower part of the radome, which doesn't affect the result.